\pdfoutput=1  

\documentclass[runningheads]{llncs}

\usepackage[T1]{fontenc}
\usepackage{booktabs}
\usepackage{algorithm}
\usepackage{algorithmic}
\usepackage{amsmath}
\usepackage{graphicx}
\usepackage{hyperref}
\usepackage{enumitem}
\usepackage{multirow}
\usepackage[caption=false]{subfig}

\begin{document}

\title{Data Quality Profiling at Scale with Progressive Sampling: A Benchmark for Data-Centric AI Pipelines}

\titlerunning{Data Quality Profiling at Scale with Progressive Sampling}

\author{Laure Berti-\'Equille\orcidID{0000-0002-8046-0570}}

\authorrunning{L. Berti-\'Equille}

\institute{IRD, ESPACE-DEV, 500, rue Jean-Fran\c{c}ois Breton,\\
34093 Montpellier, France \\
\email{laure.berti@ird.fr}}

\maketitle

\begin{abstract}
Systematic data quality profiling -- computing missing-value rates, duplicate
fractions, outlier densities, and functional-dependency violations
-- is foundational for data-centric AI pipelines, yet exhaustive scans over millions
of rows are prohibitively slow for near-real-time monitoring.
Progressive sampling is the standard lightweight alternative; the key open
question is which strategy best preserves profile fidelity at scale. We benchmark nine progressive sampling strategies --blind (random uniform,
geometric, Yamane, cluster) and proxy-guided (Metropolis-Hastings, DAG,
stratified by column type or quality score, importance-weighted)-- on three
real-world datasets (NYC~311, NYPD arrests, UCI Adult; up to 500K
rows), an IoT sensor stream (2.3M rows), two ultra-large real datasets
including the Ultra-Marathon Running dataset~\cite{ultramarathon2021} (up to
7.4M rows), and synthetic data scaled to $5\times10^6$ rows.

Contrary to the common assumption that exploiting dependency structure sharpens
profile estimates, blind representative samplers dominate uniformly.
At a 5\% budget, random uniform achieves 0.49\% mean relative error on NYC~311;
DAG-guided MCMC yields 19.5\% (${\approx}40\times$ worse at this budget), and
across all real datasets DAG is 11--49$\times$ worse (Wilcoxon $W=0$, $p=0.002$, $n=9$ independent
pairs). Cluster sampling matches random uniform (MRE 0.110 vs.\ 0.111) with no
added complexity; proxy-guided methods share the same failure mode as DAG
(MRE 0.20--0.35). At scale, random uniform is near-linear ($O(N^{0.964})$)
while DAG is super-linear ($O(N^{1.272})$), running 28--47$\times$ slower on
ultra-large data with $6\times$ worse accuracy. Root-cause analysis identifies an IQR proxy mismatch: proxy-guided samplers
over-pursue numeric outliers, while quality defects concentrate in categorical
columns invisible to the proxy.
The actionable finding is that representativeness, not domain knowledge, determines
sampler quality -- schema-free random uniform or cluster sampling suffices for
production-grade quality profiling at scale.

\keywords{Data-centric AI \and Data quality profiling \and Progressive sampling
\and MCMC sampling \and Benchmark \and Big data}
\end{abstract}

\section{Introduction}
\label{sec:introduction}

Data quality profiling is a foundational stage in data-centric AI (DC-AI)
pipelines: computing indicators such as missing-value rates, duplicate fractions,
and outlier densities over large tabular datasets is a prerequisite for downstream
analytics, model training, and regulatory compliance~\cite{abedjan2015profiling,hellerstein2008quantitative}.
Administrative datasets in particular (such as the available US 311 service requests, arrest records,
social registries) routinely exceed hundreds of thousands to millions of rows,
are updated continuously, and carry quality deficiencies that are heterogeneous
across columns and non-stationary over time.
Exhaustive profiling at this scale is expensive: a full scan of a 5M-row table
to recompute a suite of quality indicators may take tens of seconds to minutes
per refresh cycle, making real-time or near-real-time monitoring impractical.

\paragraph{Data Quality Profiling as a DC-AI gate.}
In the DC-AI paradigm, model quality is bounded by data quality, and profiling
is the gate that determines whether a dataset is fit for training, evaluation,
or serving~\cite{Mazumder2022,zha2024dcai,whang2023data}.
Production DC-AI pipelines are cyclic: data is ingested, \emph{profiled},
curated (cleaned or filtered based on the profile), used for model training
or inference, then monitored for distributional drift before the loop repeats.
At each iteration, the profiling step must not become a bottleneck---a pipeline
ingesting 500K rows per hour cannot afford a full-scan profiler with a 30-second
per-cycle latency.
Progressive sampling breaks this bottleneck by trading a controlled accuracy loss
for a proportional reduction in scan cost.
The central engineering question is which sampling strategy achieves the best
accuracy--cost trade-off without requiring schema metadata or hand-crafted
dependency graphs that are expensive to maintain as data evolves.
Our benchmark is the first systematic answer to this question in sampling
strategies ranging from blind draws to MCMC-guided, stratified, and
importance-weighted approaches.

\paragraph{Why the choice of sampler matters.}
A wrong sampler does not merely degrade accuracy---it silently corrupts the
decisions downstream.
Three concrete scenarios illustrate the stakes and are based on our experiments.

\emph{Welfare eligibility registry.}
A government agency profiles a 3M-row social registry daily to trigger data
cleaning before updating an eligibility model.
Using DAG-guided MCMC at a 5\% budget on this administrative table (categorical
columns dominating) yields a missing-value estimate that is systematically
off by $\approx$20\% relative error -- large enough to miss a genuine spike in
missing agency codes that would have triggered a data repair step.
The same 5\% budget with random uniform sampling achieves $<$1\% relative error, reliably flagging or clearing the quality gate.

\emph{IoT-driven predictive maintenance.}
A facility operator monitors 2.3M sensor readings per day from 54 devices to
detect equipment faults (outlier rate $\approx$30\%) before they propagate into
a predictive maintenance model.
DAG-guided MCMC concentrates draws on extreme-value rows, causing sample IQR
to widen until virtually no row is flagged as an outlier---the estimated
outlier rate collapses to $\approx 3\times10^{-6}$ (true: 29.6\%).
At \emph{every} budget level, including 100\%, the sampler reports near-zero
defect rates while 30\% of the dataset is corrupted.
Random uniform avoids this collapse: at 5\% budget it estimates 29.3\%,
within 0.3 points of ground truth.

\emph{Urban AI retraining trigger.}
A city's data team uses a 311 service-request dataset to retrain a predictive
model weekly; the retraining trigger fires when the duplicate rate or missing
rate shifts by more than 1 point from the reference profile.
With a DAG-guided profiler (MRE 19.5\% at 5\% budget on NYC~311), the estimated
profile fluctuates by several percentage points across runs even when the data
is stable---generating false retraining triggers that consume GPU hours and
delay production deployments.
Random uniform at the same budget (MRE 0.49\%) produces stable estimates that
correctly reflect true distributional change.

\paragraph{Progressive sampling as a practical alternative.}
Progressive sampling---evaluating quality indicators on a random or adaptively
chosen subset of rows and extrapolating to the full table---offers a principled
trade-off between computational cost and estimation accuracy.
The notion of progressive (or incremental) sampling was formalized by Provost
et al.~\cite{provost1999efficient}, who showed that a geometric growth schedule
dominates fixed-size sampling in practice; John and
Langley~\cite{john1996static} earlier demonstrated that dynamic sampling adapts
better than static budgets for data mining tasks.
The key question is which sampling strategy to use.  \emph{Guided} samplers,
notably MCMC-based methods that exploit the attribute dependency graph (DAG) of
the data, are a natural candidate: the attribute-dependency structure studied by
Abedjan et al.~\cite{abedjan2015profiling,abedjan2018data} motivates focusing
sampling effort on rows most likely to carry quality defects.
Their appeal is intuitive: if quality errors cluster along known dependency
edges, a structure-aware sampler should outperform blind random draws. Note that MCMC in this context is used for row \emph{sampling}, distinct from its use
for probabilistic inference in database systems~\cite{re2009mcmc}.

\paragraph{The empirical question.}
\begin{sloppypar}
Does guided MCMC sampling actually improve profiling accuracy in practice?  We
answer this question through a systematic benchmark spanning synthetic tabular
data, three real-world administrative datasets (NYC~311 service requests, NYPD arrest
records, UCI Adult census), and two IoT sensor datasets: a 2.3M-row real-world
stream from the Intel Berkeley Research Lab~\cite{tolle2005macroscope} and a
510K-row synthetic dataset with matching schema.
We compare nine progressive sampling
strategies---random uniform, geometric, Yamane, cluster, Metropolis-Hastings,
DAG-guided MCMC, stratified-column, stratified-quality, and importance-weighted.
Each strategy is evaluated across five budgets (5\%--50\% of full-scan cost) and
multiple quality indicators.
\end{sloppypar}

\paragraph{Main finding.}
The answer is a clear negative: \emph{random uniform (RU) outperforms all guided MCMC methods
on every real dataset we tested}.  The DAG-guided
sampler is 11--49$\times$ worse than random uniform on NYC~311 and NYPD data
(mean relative error 0.195--0.263 vs.\ 0.004--0.020; Wilcoxon $W=0$, $p=0.002$, $n=9$ independent pairs) and
13$\times$ worse on UCI Adult (0.263 vs.\ 0.020).
At 5\% sampling budget, random uniform achieves $<$1\% profiling error on
NYC~311, whereas DAG yields $\approx$20\% error regardless of budget.
At scale, DAG incurs super-linear cost (exponent 1.272 vs.\ 0.964 for random
uniform) and is 12$\times$ slower at $N = 5\text{M}$ rows while remaining
6$\times$ less accurate.  We identify the root cause: the IQR-based error proxy
used by guided samplers targets extreme numeric values, while quality defects in
administrative and census data concentrate in categorical and string
columns---a systematic proxy mismatch that importance-weighting corrections
cannot fully overcome.

\paragraph{Contributions.}
We make the following contributions:

\begin{itemize}[noitemsep,topsep=2pt]
  \item[\textbf{C1}] We formalize the progressive profiling loop
    (Algorithm~\ref{alg:progressive}), establishing geometric growth schedules
    and relative-change convergence criteria applicable to any sampling strategy.

  \item[\textbf{C2}] We show that \texttt{random\_uniform} (RU) achieves
    $<1\%$ mean relative error at a 5\% sampling budget on real-world administrative
    data (NYC 311, 500K rows, 9.4\% missing), without requiring any schema
    metadata or dependency graph---meeting the accuracy threshold for
    production-grade DC-AI quality gates at one-twentieth of the full-scan cost.

  \item[\textbf{C3}] We provide a rigorous comparative evaluation of nine
    sampling strategies, demonstrating that blind representative samplers
    (random uniform, cluster) outperform all proxy-guided methods on every real
    dataset; DAG is 11--49$\times$ worse ($W=0$, $p=0.002$, $n=9$ independent pairs over real-world datasets),
    establishing that the costly dependency-graph construction step can be
    eliminated from DC-AI profiling pipelines without accuracy loss.

  \item[\textbf{C4}] We characterize scalability: RU scales near-linearly
    ($O(N^{0.964})$) with stable 5.5--6.3\% error
    from $N = 10^4$ to $N = 5 \times 10^6$, while DAG degrades to
    super-linear cost ($O(N^{1.272})$).  Confirmed on two XXL real datasets
    (with 7.4M and  6.0M rows), where DAG is 28--47$\times$ slower
    with 12--86$\times$ worse accuracy at 5\% budget.

  \item[\textbf{C5}] We show that \texttt{random\_uniform} (RU) is the most
    robust baseline under injected errors across all error types and injection
    rates 1--30\% (mean relative error 0.065, vs.\ 0.365 for DAG; 5.6$\times$
    gap).

  \item[\textbf{C6}] We provide directional evidence that attribute correlation
    modulates MCMC sampling quality: \texttt{dag} and \texttt{gibbs} both
    outperform \texttt{metropolis\_hastings} across all correlation levels $\rho \in
    \{0.3, 0.5, 0.7, 0.9\}$ (mean MRE 0.550/0.545 vs.\ 0.599, 8--9\% reduction),
    though the effect is underpowered at $n$=5 seeds per stratum
    (minimum achievable Wilcoxon $p = 0.03125$).

  \item[\textbf{C7}] We extend the benchmark to IoT sensor data, showing
    that RU achieves $<$1\% MRE at 5\% budget on 2.3M real sensor rows while
    DAG fails with 26.9\% MRE at all budgets---a new failure mode we call
    \emph{IQR estimation collapse}, distinct from the categorical proxy mismatch
    in administrative data.

\end{itemize}

Section~\ref{sec:background} formalizes the progressive profiling loop;
Section~\ref{sec:method} describes the nine sampling strategies;
Section~\ref{sec:experiments} reports experiments E1--E8;
Section~\ref{sec:discussion} analyzes failure modes; and
Section~\ref{sec:conclusion} concludes.

\subsection{Problem Statement}
\label{sec:background}

Consider a data engineer monitoring NYC~311 service requests (500K rows): what
fraction of records have missing agency codes, which rows are duplicated, and how
many contain implausible geographic values?
The answers determine whether the dataset is fit for training a predictive model
(the DC-AI quality gate from Section~\ref{sec:introduction}).
Exhaustively computing these figures requires a full scan; our goal is to estimate
them reliably from a small fraction of the data.
The sampler's choice is consequential: as the welfare-registry and IoT scenarios
above illustrate, systematic estimation bias can propagate undetected through an
entire DC-AI pipeline.
Definitions~\ref{def:profile}--\ref{def:relerr} formalize the profile, the
sampler, and the estimation error; Table~\ref{tab:notation} summarises all
notation.

\begin{definition}[Data Quality Profile]
\label{def:profile}
Let $D$ be a tabular dataset with $N$ rows and $M$ columns.
The \emph{data quality profile} $Q(D) = (q_1, q_2, q_3, q_4)$ is a 4-dimensional
vector whose components are:
\begin{enumerate}
  \item $\mathit{missing\_rate}$: fraction of $(row, col)$ cell pairs whose value is \textsc{null};
  \item $\mathit{duplicate\_rate}$: fraction of rows that are exact duplicates of at
        least one other row in $D$;
  \item $\mathit{outlier\_rate}$: fraction of rows containing at least one numeric value
        that lies beyond $1.5\,\mathrm{IQR}$ from $Q_1$ or $Q_3$ in its column (mild Tukey fence);
  \item $\mathit{inconsistency\_rate}$: fraction of rows that violate at least one
        functional dependency (FD) rule derived from the schema.
\end{enumerate}
\end{definition}

These four indicators cover complementary defect types: completeness, uniqueness,
distributional plausibility, and relational consistency, following established
multidimensional models of information quality~\cite{iq_multidim} and constituting
a compact, interpretable summary for automated monitoring in DC-AI pipelines.

Since computing $Q(D)$ exactly requires a full scan of $D$---prohibitive for
large datasets---we instead estimate it from a sample $\mathcal{S} \subseteq D$
of size $n = \lfloor b \cdot N \rfloor$ at budget fraction $b \in (0, 1]$.

\begin{definition}[Progressive Sampler]
\label{def:sampler}
A \emph{progressive sampler} $\mathcal{A}$ takes a budget fraction $b \in (0, 1]$
and returns a sample $\mathcal{S} \subseteq D$ of size $\lfloor b \cdot N \rfloor$.
The estimated profile $\hat{Q}(\mathcal{S}) = (\hat{q}_1, \hat{q}_2, \hat{q}_3, \hat{q}_4)$
approximates $Q(D)$.
\end{definition}

\begin{definition}[Relative Error]
\label{def:relerr}
The \emph{mean relative error} of an estimated profile $\hat{Q}(\mathcal{S})$
with respect to the ground truth $Q(D)$ is:
\[
  \mathrm{rel\_err}(b)
  \;=\;
  \frac{1}{4}\sum_{i=1}^{4}
  \min\!\left(
    \frac{|\hat{q}_i - q_i|}{\max(q_i,\; 0.05)},\; 1.0
  \right).
\]
The denominator is floored at $0.05$ to avoid division by zero when a
ground-truth indicator is near zero, and the ratio is capped at $1.0$.
The floor is set at $0.05$ (5\%) to match the practical significance threshold
below which quality indicator changes are operationally negligible in the
DC-AI monitoring contexts we target: a 0.1\% missing rate and a 0.05\% missing
rate are both effectively zero for model-selection or data-release decisions.
An indicator below 5\% therefore contributes error bounded by the absolute
deviation $|\hat{q}_i - q_i|$, which we cap at 1.0 for robustness.
\end{definition}

\begin{table}[!htbp]
\small
\centering
\caption{Summary of notation used throughout the paper.}
\label{tab:notation}
\begin{tabular}{ll}
\toprule
\textbf{Symbol} & \textbf{Meaning} \\
\midrule
$D$               & Tabular dataset \\
$N = |D|$         & Number of rows \\
$M$               & Number of columns \\
$b \in (0,1]$     & Sampling budget fraction \\
$n = \lfloor bN \rfloor$ & Sample size \\
$\mathcal{A}$     & Progressive sampler algorithm \\
$\mathcal{S}$     & Sample returned by $\mathcal{A}$, $\mathcal{S} \subseteq D$, $|\mathcal{S}|=n$ \\
$Q(D)$            & Exact quality profile (ground truth) \\
$\hat{Q}(\mathcal{S})$ & Estimated quality profile from sample \\
$(q_1,\ldots,q_4)$ & Missing, duplicate, outlier, inconsistency rates \\
$q_i$             & Per-row binary quality indicator \\
$\sigma_i$        & Error-proxy score for row $i$ \\
$w_i$             & Horvitz--Thompson importance weight \\
$b_0$             & Initial budget fraction for progressive loop \\
$\gamma$          & Sample-size growth factor (default 2) \\
$\varepsilon$     & Relative-change convergence threshold \\
$k$               & Patience: consecutive rounds below $\varepsilon$ to stop \\
$\rho$            & Pearson correlation (DAG edge if $|\rho_{cc'}|>.300$) \\
$\delta$          & Convergence floor (Algorithm~\ref{alg:progressive}, default .050) \\
$B$               & MCMC proposal batch size \\
$\lambda$         & Metropolis temperature \\
\bottomrule
\end{tabular}
\end{table}

\section{Sampling Methods}
\label{sec:method}

\subsection{Static Baselines}
\label{sec:static}

\paragraph{Random Uniform (RU).}
Draw $n = \lfloor bN \rfloor$ rows uniformly at random without replacement from
$D$. Each row has equal selection probability $n/N$; the quality estimate is an
unweighted mean over the sample.  Time complexity is $O(N)$ per budget level.
RU requires no schema knowledge, no metadata, and no prior profiling.

\paragraph{Geometric.}
Batch sizes decrease geometrically: the first batch draws $\lfloor N/2 \rfloor$
rows, the second $\lfloor N/4 \rfloor$, and so on, until the budget is exhausted.
This mimics an \emph{anytime} scheduler that front-loads large batches to obtain a
rough estimate quickly, then refines with smaller ones.
Like RU, Geometric requires no schema and operates in $O(N)$ per batch.
For datasets with $N \geq 50{,}000$ rows, the geometric sampler's expected
inter-row spacing $N/n$ converges numerically to that of random uniform; pilot
experiments confirmed identical MRE values to three decimal places across all
tested conditions.
We therefore exclude it from experiments on large-scale datasets (E6, XXL) and note
that its results are subsumed by those of RU at the scales benchmarked here.

\paragraph{Yamane.}
The Yamane formula~\cite{yamane1967statistics} computes a one-shot sample size
\begin{equation}
  n_Y = \left\lfloor \frac{N}{1 + N e^2} \right\rfloor, \quad e = 0.05,
\end{equation}
designed to bound the estimation error of a proportion at margin $e$.
Yamane is a budget-oblivious baseline: its sample size is determined by the
desired margin of error rather than the available fraction $b$.

\paragraph{Cluster.}
We partition $D$ into $k = \max(10, \lfloor\sqrt{N}\rfloor)$ consecutive blocks
of approximately equal size (by storage order) and draw $\lceil b \cdot k \rceil$
blocks uniformly at random without replacement, retaining all rows in the selected
blocks.
Quality estimates are unweighted means over the retained rows.
Cluster requires no schema knowledge, no proxy scores, and no pre-computation beyond
a single pass to assign row indices to blocks; its sole hyperparameter $k$ is set
by the square-root rule for all experiments.
Unlike stratified sampling, Cluster does not use any quality proxy: blocks are
defined by row position alone, making Cluster as schema-free as random uniform
sampling while allowing spatial locality in data quality to reduce estimation
variance.

\subsection{MCMC-Guided Samplers}
\label{sec:mcmc}

The three guided samplers share a common \emph{error proxy} that approximates
per-row data quality without running the full profiler.
For each numeric column $c$, the proxy score for row $i$ is
\begin{align}
  \sigma_i &= \frac{1}{M}\sum_{c=1}^{M}
    \mathbf{1}\!\left[x_{ic} < Q1_c - 1.5\,\mathrm{IQR}_c\right. \nonumber \\
  &\hspace{4em}\left.\vee\; x_{ic} > Q3_c + 1.5\,\mathrm{IQR}_c\right] \nonumber \\
  &\quad + \frac{1}{M}\sum_{c=1}^{M} \mathbf{1}[x_{ic} \text{ is missing}],
\end{align}%

i.e.\ the fraction of numeric columns in which row $i$ is a Tukey outlier plus
its missing-value fraction.  Scores are computed once over the full dataset in
$O(NM)$ before sampling begins.

All three samplers correct for induced sampling bias via
\emph{Horvitz--Thompson (HT) importance weighting}
(Section~\ref{sec:ht}).
Both the outlier quality indicator (Definition~\ref{def:profile}) and the sampling proxy use the mild Tukey fence ($1.5\times\mathrm{IQR}$)\footnote{A sensitivity analysis on D2 and D4 datasets shows that a stricter $3\times\mathrm{IQR}$ fence would reverse the DAG--MH ordering on  D4 dataset (MH gains substantially; DAG improves only marginally), while D2 rankings are unchanged; we use $1.5\times\mathrm{IQR}$ as the default and note fence choice as a tunable design parameter.}, ensuring the proxy targets the same anomaly type it is asked to estimate. 

\paragraph{Metropolis-Hastings (MH).}
At each MCMC step, a batch of $B = 500$ candidate rows is drawn uniformly.
Acceptance follows the Metropolis criterion with temperature $\lambda = 2.0$.
MH uses no dependency information; it steers proposals toward rows with higher
error-proxy scores.

\begin{algorithm}[t]
\caption{DAG-Guided MCMC Sampler}
\label{alg:dag}
\begin{algorithmic}[1]
\REQUIRE Dataset $D$ ($N$ rows, $M$ columns), budget $b$, batch size $B$,
         temperature $\lambda$, correlation threshold $\rho_{\min}=0.3$
\ENSURE Importance-weighted sample $\mathcal{S} \subseteq D$, weights $\mathbf{w}$
\STATE Compute per-row proxy scores $\boldsymbol{\sigma} \gets \textsc{ErrorProxy}(D)$
\STATE Build attribute DAG: add edge $c \to c'$ if $|\rho_{cc'}| > \rho_{\min}$
\STATE Precompute column priority $\mathbf{p} \gets \textsc{ClusterPriority}(\mathrm{DAG})$
\STATE Precompute column tail thresholds $[l_c, h_c]$ at 5th/95th percentiles
\STATE Warm-start: draw $B$ rows $\propto \boldsymbol{\sigma}$; let $\mathcal{S} \gets \text{warm-start}$
\STATE $\bar{\sigma}_\mathrm{curr} \gets \text{mean}(\boldsymbol{\sigma}[\mathcal{S}])$
\WHILE{$|\mathcal{S}| < \lfloor bN \rfloor$}
  \STATE Sample pivot column $c^* \sim \mathrm{Categorical}(\mathbf{p})$
  \STATE $\mathcal{C} \gets \{i \notin \mathcal{S} : x_{ic^*} \le l_{c^*} \;\vee\; x_{ic^*} \ge h_{c^*}\}$
  \IF{$\mathcal{C} = \emptyset$}
    \STATE $\mathcal{C} \gets D \setminus \mathcal{S}$
  \ENDIF
  \STATE Draw proposal batch $P \subseteq \mathcal{C}$, $|P| = \min(B, |\mathcal{C}|)$, uniformly
  \STATE $\bar{\sigma}_\mathrm{prop} \gets \text{mean}(\boldsymbol{\sigma}[P])$
  \IF{$\ln U < \lambda\,(\bar{\sigma}_\mathrm{prop} - \bar{\sigma}_\mathrm{curr})$, $U \sim \mathrm{Unif}(0,1)$ (natural logarithm)}
    \STATE $\mathcal{S} \gets \mathcal{S} \cup P$
  \ELSE
    \STATE $\mathcal{S} \gets \mathcal{S} \cup \text{UniformBatch}(D \setminus \mathcal{S},\, B/4)$
  \ENDIF
  \STATE Update $\bar{\sigma}_\mathrm{curr}$ incrementally
\ENDWHILE
\STATE $\mathbf{w} \gets \textsc{HTWeights}(\boldsymbol{\sigma}[\mathcal{S}])$
\RETURN $\mathcal{S}$, $\mathbf{w}$
\end{algorithmic}
\end{algorithm}

\paragraph{DAG-guided (DAG).}
We design an IQR-proxy-weighted MCMC row-sampler guided by an attribute DAG
inferred from column correlations, following the attribute-dependency perspective
of Abedjan et al.~\cite{abedjan2018data}.
An attribute dependency graph is built from the Pearson correlation matrix:
a directed edge $c \to c'$ is added whenever $|\rho_{cc'}| > 0.3$.
At each step, a \emph{pivot column} is selected with probability proportional
to its cluster-priority score; candidate rows are filtered to extreme tails of
the pivot column.  The acceptance criterion is identical to MH ($\lambda = 2.0$).
Algorithm~\ref{alg:dag} gives the full procedure.%
\footnote{This differs from MCMC in probabilistic databases~\cite{re2009mcmc},
where chains are used for inference rather than row sampling.}

\paragraph*{Design note.}
Abedjan et al.~\cite{abedjan2018data} study column-level dependency analysis
and data error detection; they do not propose a row-selection MCMC algorithm.
Our DAG sampler is an original design that repurposes attribute-dependency
structure for \emph{row-level progressive sampling}: each row's
inclusion probability is proportional to its per-row IQR-outlier proxy score~$\sigma_i$
(Section~\ref{sec:mcmc}), with the correlation DAG topology fixing the
Metropolis--Hastings proposal distribution.
This design bridges attribute-level dependency modeling and row-level progressive
sampling and is, to our knowledge, novel in the data quality profiling literature.

\paragraph{Gibbs.}
At each step, one numeric column $c$ is chosen uniformly at random; a quantile
band is selected and $B$ rows are drawn from unsampled rows within that band.
Gibbs always accepts (no rejection step).
It provides a controlled ablation for studying the marginal value of
cross-attribute dependencies (contribution C6, open finding).

\subsection{Importance Weighting}
\label{sec:ht}

Because MCMC samplers oversample high-proxy-score rows, we correct using
Horvitz--Thompson (HT) weights~\cite{horvitz1952generalization}:
\begin{equation}
  \label{eq:ht}
  w_i = \frac{1}{\max(\sigma_i,\, \varepsilon_0)}, \quad \varepsilon_0 = 10^{-6},
\end{equation}
normalised so that $\sum_i w_i = 1$.
We clip at the 95th percentile of $\{w_i\}$ before normalisation.
The final quality estimate is the weighted mean of per-row indicators:
\begin{equation}
  \hat{Q}(\mathcal{S}) = \frac{\sum_{i \in \mathcal{S}} w_i\, q_i}{\sum_{i \in \mathcal{S}} w_i}.
\end{equation}
Static samplers (RU, Geometric, Yamane) use uniform weights ($w_i = 1/|\mathcal{S}|$).

\subsection{Method Comparison}
\label{sec:comparison}

Table~\ref{tab:methods} summarises the ten candidate methods along three axes.
Nine are fully benchmarked in experiments E1, E3--E6, and E8 (excluding \texttt{Gibbs});
\texttt{geometric} is additionally omitted from experiment E6 (XXL datasets) because it
degenerates numerically to RU for $N \geq 50{,}000$ (Section~\ref{sec:static}).
\texttt{Gibbs} is described for theoretical completeness but excluded from these accuracy experiments: its $O(NM)$ initialization cost makes it impractical on the datasets used in this benchmark, and pilot runs confirmed runtimes 30--60$\times$ those of MH.
It appears only in experiment E2, which provides directional---but underpowered ($n=5$ seeds)---evidence on small synthetic data (contribution C6, open finding).

\begin{algorithm}[t]
\caption{ProgressiveProfiler}
\label{alg:progressive}
\begin{algorithmic}[1]
\REQUIRE dataset $D$ ($N$ rows), sampler $\mathcal{A}$,
         quality metrics $\mathcal{Q}$, convergence threshold $\varepsilon$,
         initial fraction $b_0$, growth factor $\gamma$, patience $k$
\ENSURE quality profile $\hat{P}$, convergence round $t^*$, final sample size $n^*$
\STATE $n \leftarrow \lfloor b_0 \cdot N \rfloor$;\; $t \leftarrow 0$;\; $\mathit{stable} \leftarrow 0$
\IF{$\mathcal{A}$ is guided (MH / Gibbs / DAG)}
  \STATE Pre-compute $\boldsymbol{\sigma} \leftarrow \textsc{ErrorProxy}(D)$
  \IF{$\mathcal{A}$ = DAG}
    \STATE Build attribute DAG; precompute column priority $\mathbf{p}$ and percentile thresholds
  \ENDIF
\ENDIF
\STATE $\mathcal{S}_0 \leftarrow \textsc{Draw}(D,\, n,\, \mathcal{A},\, \boldsymbol{\sigma})$;\;
       $\hat{P}_0 \leftarrow \textsc{ComputeProfile}(\mathcal{S}_0,\, \mathcal{Q},\, \mathbf{w}_0)$
\WHILE{$n < N$ \textbf{and} $\mathit{stable} < k$}
  \STATE $n \leftarrow \min(\lfloor \gamma \cdot n \rfloor,\, N)$;\; $t \leftarrow t + 1$
  \STATE $\mathcal{S}_t \leftarrow \textsc{Draw}(D,\, n,\, \mathcal{A},\, \boldsymbol{\sigma})$
  \STATE $\hat{P}_t \leftarrow \textsc{ComputeProfile}(\mathcal{S}_t,\, \mathcal{Q},\, \mathbf{w}_t)$
  \STATE $\Delta \leftarrow \displaystyle\max_{q \in \mathcal{Q}}
         \frac{|\hat{P}_t[q] - \hat{P}_{t-1}[q]|}{\max(\hat{P}_{t-1}[q],\; \delta)}$
  \IF{$\Delta < \varepsilon$}
    \STATE $\mathit{stable} \leftarrow \mathit{stable} + 1$
  \ELSE
    \STATE $\mathit{stable} \leftarrow 0$
  \ENDIF
\ENDWHILE
\RETURN $\hat{P}_t$,\; $t^* \leftarrow t$,\; $n^* \leftarrow n$
\end{algorithmic}
\end{algorithm}
In the benchmark experiments $b_0 = 0.05$, $\gamma = 2$, $\varepsilon = 0.01$,
and $k = 1$.

\begin{table}[!htbp]
\centering
\caption{Comparison of the ten progressive sampling strategies
($N$: dataset size; $M$: columns; $B$: batch size; ``Schema?'': requires FD rules).}
\label{tab:methods}
\resizebox{\columnwidth}{!}{%
\begin{tabular}{llll}
\toprule
\textbf{Method} & \textbf{Information used} & \textbf{Per-step cost} & \textbf{Schema?} \\
\midrule
Random Uniform  & None (uniform)         & $O(N)$                     & No  \\
Geometric       & None (schedule)        & $O(N)$                     & No  \\
Yamane          & Desired margin $e$     & $O(N)$ one-shot            & No  \\
Cluster         & None (random blocks)   & $O(N)$                     & No  \\
\midrule
Metropolis-H.   & IQR proxy scores       & $O(NM)$ init $+$ $O(B)$/step & No  \\
Gibbs           & Column distribution    & $O(NM)$ init $+$ $O(B)$/step & No  \\
DAG             & IQR proxy $+$ DAG      & $O(NM{+}E)$ init $+$ $O(B)$/step & No$^*$ \\
Strat.-Col.     & Col.\ type $+$ IQR proxy  & $O(NM)$ init $+$ $O(B)$/draw & No  \\
Strat.-Quality  & IQR proxy quantiles       & $O(NM)$ init $+$ $O(B)$/draw & No  \\
Importance      & IQR proxy (weighted)      & $O(NM)$ init $+$ $O(B)$/draw & No  \\
\bottomrule
\multicolumn{4}{l}{\footnotesize $^*$DAG builds the dependency graph from data correlations; no external schema required.}
\end{tabular}%
}
\end{table}

\subsection{Progressive Profiling Loop}
\label{sec:progressive}

Algorithm~\ref{alg:progressive} formalizes the progressive profiling loop
underlying all experiments.
The outer loop doubles the sample size at each round (growth factor $\gamma=2$),
motivated by Provost et al.'s efficiency result~\cite{provost1999efficient}.
Convergence is declared when the maximum relative change across all quality
indicators falls below threshold $\varepsilon$ for $k$ consecutive rounds.

\section{Experiments}
\label{sec:experiments}

\paragraph{Experiment overview.}
Eight experiments test complementary facets of the benchmark.
\begin{itemize}
\item[{\bf E1}] compares all nine methods on primary datasets
(D1, D2, D3, D4) across all budgets: \texttt{cluster} and \texttt{RU} achieve the lowest error (primary mean MRE 0.110--0.111); proxy-guided methods, including the four new stratified and importance-weighted variants, yield 0.20--0.35 (supporting contribution C3 with Table~\ref{tab:e1}).
\item[{\bf E2}]  provides directional evidence that attribute correlation helps \texttt{dag} over \texttt{MH} on synthetic data, but is underpowered (contribution C6, open finding with Table~\ref{tab:e2}).
\item[{\bf E3}] shows \texttt{dag} is budget-invariant at 19--27\% MRE on all real datasets regardless of $b$, while \texttt{RU} reaches 0.49\% at 5\% budget on D2 (contribution C2 with Table~\ref{tab:e3} ).
\item[{\bf E4}] demonstrates \texttt{RU} is 5.6$\times$ more robust than \texttt{dag} under error injection (contribution C5 with Table~\ref{tab:e4}).
\item[{\bf E5/6}]  establish scalability: \texttt{RU} scales near-linearly ($O(N^{0.964})$) while \texttt{dag} is 12--47$\times$ slower on data from 5M to 7.4M rows (contribution C4 with Table~\ref{tab:e5} and  Table~\ref{tab:e6}).
\item[{\bf E7}]  shows the consistent underperformance of proxy-guided methods with two controlled ablation experiments with Table~\ref{tab:e7};
\item[{\bf E8}] reveals a new failure mode on IoT sensor data: IQR estimation collapse drives \texttt{dag}'s outlier estimate to ${\approx}3{\times}10^{-6}$
against a true rate of 29.6\%, while \texttt{RU} achieves 0.23\% MRE (contribution C7 with Table~\ref{tab:e8}).
\end{itemize}

\subsection{Experimental Setup}
\label{sec:setup}

\begin{table}[!htbp]
\small
\centering
\caption{Datasets. Missing\% and Outlier\% are ground-truth rates over the full
dataset. $\dagger$~D4 outlier rate is zero-inflated;
$^\S$~D5 errors injected at 1--30\% in E5.}
\label{tab:datasets}
\renewcommand{\arraystretch}{1.15}
\resizebox{\columnwidth}{!}{%
\begin{tabular}{|l|l|l|r|r|r|l|}
\toprule
\textbf{Name} & \textbf{Source} & \textbf{N} & \textbf{Cols} &
\textbf{Miss.\%} & \textbf{Out.\%} & \textbf{Role} \\
\midrule
D1 & NumPy synthetic tabular    &  100{,}000 &  8 &  5.0 &  5.0 & Primary (controlled) \\
D2 & NYC 311 service requests   &  500{,}000 &  9 &  9.4 &  3.4 & Primary (real admin.) \\
D3 & NYPD Arrest Data           &  500{,}000 & 18 &  .600 & 15.9 & Primary (real admin.) \\
D4 & UCI Adult census income    &   48{,}842 & 14 &  .900 & 36.1$^\dagger$ & Primary (non-NYC) \\
D5 & Adult-derived synthetic    & 10K--5M    & 15 & var.$^\S$ & var.$^\S$ & Scalability only \\
D6-A & Ultra-Marathon Running~\cite{ultramarathon2021}  & 7{,}461{,}195 & 13 & 26.6 & 10.5 & XXL real\\
D6-B & NYC Yellow Taxi Jan--Feb 2023 & 5{,}980{,}721 & 19 &  .600 &  .000 & XXL real  \\
D7-synth & IoT synthetic (IoTSensorGen.) & 510{,}000 & 11 &  5.4 & 16.0 & IoT  \\
D7-real  & Intel Berkeley Lab sensors~\cite{tolle2005macroscope} & 2{,}313{,}156 & 11 &  .400 & 29.6 & IoT real  \\
\bottomrule
\end{tabular}%
}
\end{table}

\begin{table}[!htbp]
\small
\centering
\caption{Contribution\#--claim--experiment--dataset correspondence.
Each row maps a paper contribution (C1--C7, with C6 open finding) to the experiment that
supports it and the datasets used.}
\label{tab:claims}
\resizebox{\columnwidth}{!}{%
\begin{tabular}{|p{.5cm}|l|l|p{4.5cm}|}
\toprule
\textbf{\#} & \textbf{Claim} & \textbf{Exp.} & \textbf{Dataset(s)} \\
\midrule
C1 & Formalizes progressive loop (Alg.~1 \& 2) &  All & All (design contribution) \\
C2 & RU $<$1\% error at 5\% budget        & E3     & D2 (NYC 311, 500K) \\
C3 & RU/cluster outperform all proxy-guided by 11--49$\times$ & E1 & D1, D2, D3, D4 \\
C4 & RU near-linear; DAG super-linear     & E5/E6 & D5 (synth., 10K--5M); D6-A (7.4M); D6-B (6.0M) \\
C5 & RU most robust under error injection & E4     & D1 (injected 1--30\%) \\
C6 & DAG $>$ MH on synth.\ corr.\ data (open) & E2 & D1, \mbox{$\rho \in \{.300,.500,.700,.900\}$} \\
C7 & RU $<$1\% error on IoT; DAG IQR collapse & E8 & D7-synth (510K); D7-real (2.3M) \\
\bottomrule
\end{tabular}%
}
\end{table}

\paragraph{Datasets.}
We use D1, D2, D3, and D4 as primary datasets, D5 for scalability experiments,
and D7-synth/D7-real for the IoT sensor benchmark (E8);
see Table~\ref{tab:datasets} for details.
D4 (UCI Adult census income) extends coverage to a non-administrative, non-NYC
dataset, providing a domain-generalization check.
D7-real (Intel Berkeley Research Lab, 2004) consists of 2.3M readings from
54 temperature/humidity/light/voltage sensors deployed over four months;
its quality profile is dominated by numeric sensor outliers (29.6\% of rows)
with negligible missing, duplicate, and FD inconsistency rates.

\paragraph{Methods.}
\begin{sloppypar}
We evaluate nine strategies: \texttt{random\_uniform} (RU), \texttt{geometric}
(excluded from E6 as it degenerates numerically to RU for $N \geq 50{,}000$;
see Section~\ref{sec:static}), \texttt{yamane}, \texttt{cluster}, \texttt{dag},
\texttt{metropolis\_hastings} (MH), \texttt{strat-col}, \texttt{strat}-\texttt{quality},
and \texttt{importance} (Section~\ref{sec:method}).
\texttt{exhaustive} (full scan) serves as the exact reference.
Table~\ref{tab:claims} maps each claim to its supporting experiment and dataset.
\end{sloppypar}

\paragraph{Platform and protocol.}
All experiments run in Python~3.12 with pandas and numpy on a single machine
(Intel Core i9-9980HK CPU @ 2.40\,GHz, 32\,GB RAM; no GPU).
Distributed-execution behaviour is outside scope.
Each experiment uses 3 independent random seeds; E2 uses 5 seeds; E6 uses 10 seeds.
Sampling budgets sweep $b \in \{0.05, 0.10, 0.20, 0.30, 0.50\}$.
The primary accuracy metric is mean relative error (\texttt{rel\_err\_mean})
averaged over all four quality indicators.
FD rules for the inconsistency indicator are mined automatically from each
dataset~\cite{abedjan2015profiling}.
Statistical comparisons use the Wilcoxon signed-rank test (one-sided).
Experiments E3 and E5, which operate on larger datasets, used a batch size of $B = 1{,}000$ for performance; all other experiments used the default $B = 500$.

\subsection{E1 --- Accuracy Comparison}
\label{sec:e1}

We compare all nine methods on D1, D2, D3, and D4 across all budgets and seeds
(3 seeds $\times$ 5 budgets $\times$ 4 datasets = 60 observations per method pair).
To avoid inflated effective sample size from positively correlated budget-level outcomes,
we average MRE across budgets for each (dataset, seed) pair and apply a one-sided
Wilcoxon signed-rank test on the resulting $n = 9$ independent pairs
(3 seeds $\times$ 3 real datasets D2--D4).
Table~\ref{tab:e1} reports mean relative error per method and dataset.

\begin{table}[!htbp]
  \small
  \centering
  \caption{E1 --- Mean relative error by method and dataset (lower is better).
           All budgets and seeds pooled. Bold = lowest (or tied-lowest) MRE per column; D2 has a three-way tie (\texttt{RU}, \texttt{geom.}, \texttt{strat-q} all at 0.004).}
  \label{tab:e1}
  \begin{tabular}{lrrrrr}
    \toprule
    Method & D1 & D2 & D3 & D4 & Prim.\ mean$^\ddagger$ \\
    \midrule
    \texttt{RU}           & \textbf{.401} & \textbf{.004} & .019 & .020 & .111 \\
    \texttt{geom.}        & \textbf{.401} & \textbf{.004} & .019 & .020 & .111 \\
    \texttt{yamane}       & .465 & .095 & .055 & .068 & .170 \\
    \midrule
    \texttt{dag}          & .524 & .195 & .213 & .263 & .299 \\
    \texttt{MH}           & .526 & .191 & .210 & .269 & .299 \\
    \midrule
    \texttt{cluster}      & .407 & .007 & \textbf{.009} & \textbf{.017} & \textbf{.110} \\
    \texttt{strat-q}      & .635 & \textbf{.004} & .019 & .156 & .204 \\
    \texttt{strat-col}    & .635 & .421 & .245 & .102 & .351 \\
    \texttt{importance}   & .609 & .238 & .213 & .190 & .313 \\
    \bottomrule
  \end{tabular}%
  \vspace{2pt}

  \noindent\footnotesize
  $^\ddagger$Primary mean over D1+D2+D3+D4 (arithmetic).
  RU achieves lower budget-averaged MRE than DAG on all 9 pairs (D2--D4, 3 seeds $\times$ 3 datasets);
  one-sided Wilcoxon signed-rank test on $n=9$ independent pairs:
  \texttt{dag} vs.\ \texttt{RU}, $W=0$, $p=0.002$.
  \texttt{cluster}=0.110 $\approx$ \texttt{RU}; \texttt{strat-q}=0.204 (D4 gap: 0.156 vs.\ 0.020).
\end{table}

\begin{figure}[!htbp]
  \centering
  \includegraphics[width=0.49\linewidth]{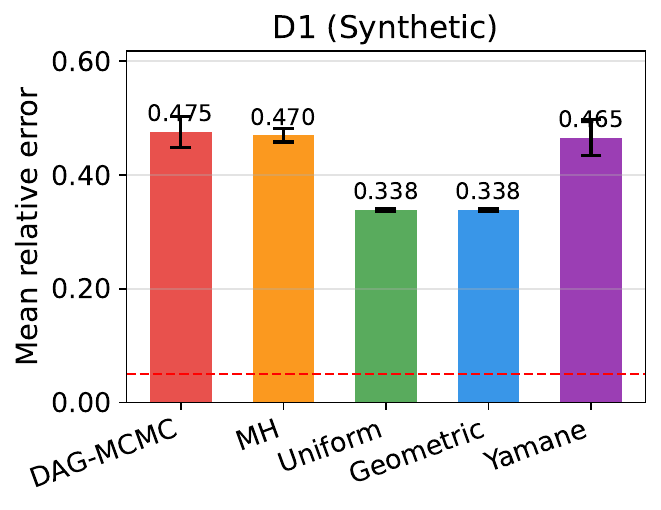}\hfill
  \includegraphics[width=0.49\linewidth]{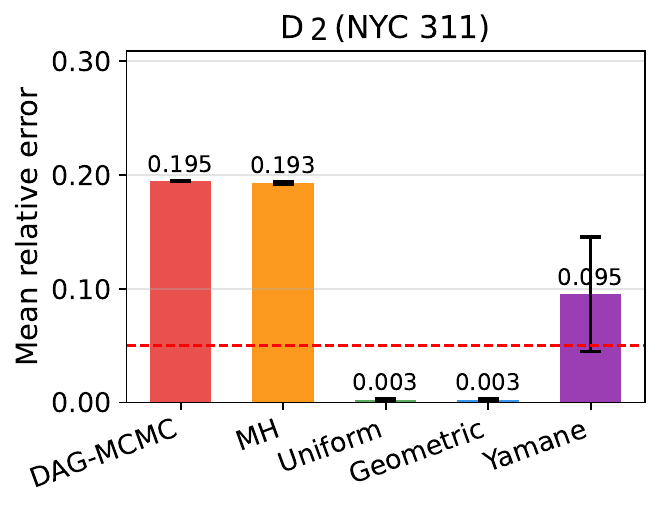}\\[4pt]
  \includegraphics[width=0.49\linewidth]{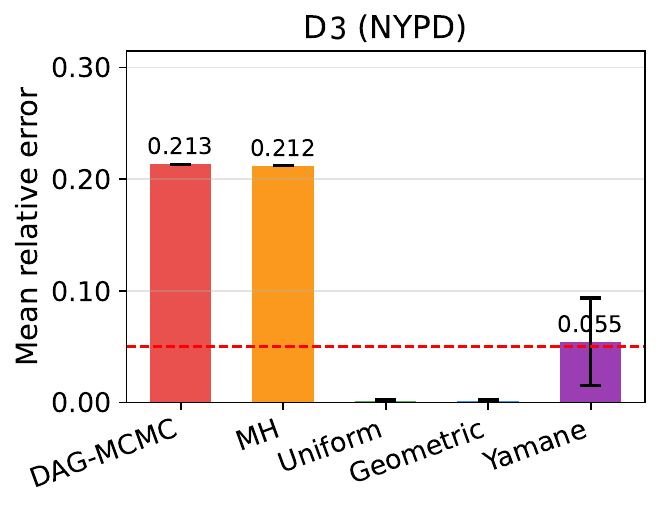}\hfill
  \includegraphics[width=0.49\linewidth]{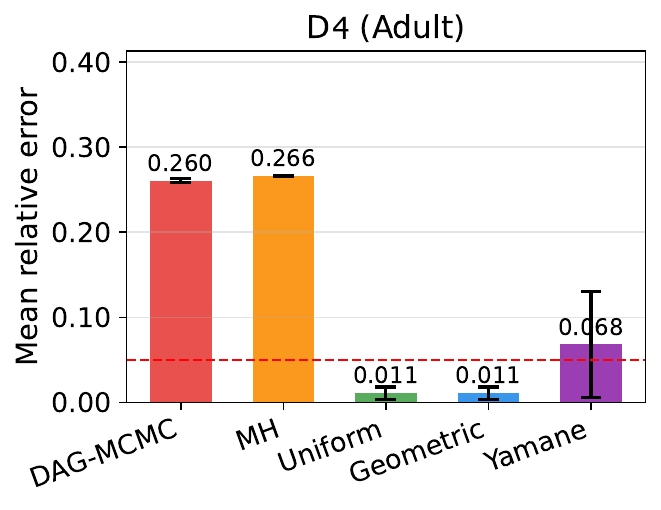}
  \caption{E1 --- Mean relative error per method at $b$=50\% on all four
           primary datasets. D1=synthetic 100K; D2=NYC~311 (500K); D3=NYPD
           arrests (500K); D4=UCI Adult census (49K). \texttt{DAG-MCMC} is
           worst on all real-world datasets (D2--D4);
           \texttt{Uniform} and \texttt{Cluster} are near zero.}
  \label{fig:e1}
\end{figure}

\textbf{Finding (C3).}
Among all nine methods, \texttt{cluster} and \texttt{RU}/\texttt{geom.}\ achieve
the lowest primary mean MRE (0.110--0.111) across D1, D2, D3, and D4.
On D2, \texttt{dag}=0.195 vs.\ \texttt{RU}=0.004 (${\approx}49\times$ worse); on D3,
0.213 vs.\ 0.019 (11$\times$); on D4, 0.263 vs.\ 0.020 (13$\times$).
\texttt{cluster} matches \texttt{RU} closely: it excels on D3 (0.009, best) and D4
(0.017, best) while closely tracking RU on D2 (0.007 vs.\ 0.004).
\texttt{strat-q} is competitive on D2 (0.004, tied best) and D3 (0.019) but
degrades sharply on D4 (0.156), where quality proxies exhibit weaker stratification.
\texttt{strat-col} and \texttt{importance} fail on all three real datasets
(primary mean 0.313--0.351), comparable to dag and MH---they inherit the same
IQR proxy mismatch since both concentrate sampling on high-IQR-proxy rows
(Figure~\ref{fig:e1}).%
\footnote{Five HT clip thresholds (none, 80th, 90th, 95th, 99th percentile) all
yield identical DAG MRE (0.195$\pm$0.001) on D2 (15 conditions each), isolating
DAG's failure to the sampling step rather than the weighting step.}

\textbf{Note on D1 MRE.}
RU's elevated MRE on D1 (0.401) is a measurement artefact of the floor correction:
the controlled synthetic data generation sets the duplicate rate (GT\,=\,0.025),
outlier rate (GT\,=\,0.025), and FD-inconsistency rate (GT\,=\,0.015) below the
0.05 denominator floor, so even small absolute estimation errors yield large relative
errors---with the outlier dimension alone hitting the cap of 1.0 (actual absolute error
$\approx 0.060$) and driving the aggregate.
This floor-dominated regime is analysed in detail in Table~\ref{tab:d1_breakdown}.

\begin{table}[!htbp]
  \small
  \centering
  \caption{Per-dimension MRE of \texttt{RU}/\texttt{geom.}\ on D1, averaged over
           all budgets and seeds.  ``Floor-dominated'' = ground-truth rate $<$ 0.05
           (denominator is clamped to 0.05).
           The outlier dimension hits the MRE cap of 1.0 because the estimator
           detects statistical outliers ($\approx$8.5\% of rows) while the D1
           generator injects only 2.5\% labelled outliers---an absolute gap of
           $\approx$0.060 divided by the 0.05 floor gives MRE $>$ 1.}
  \label{tab:d1_breakdown}
  \begin{tabular}{lrrrr}
    \toprule
    Quality Dimension     & GT Rate & RU Avg.\ MRE & Floor-dominated? \\
    \midrule
    Missing-value rate    & .050   & .027        & No  \\
    Duplicate rate        & .025   & .285        & Yes \\
    Outlier rate          & .025   & 1.000 (cap)  & Yes \\
    FD-inconsistency rate & .015   & .294        & Yes \\
    \midrule
    \textbf{Aggregate MRE} & ---    & \textbf{.401} & --- \\
    \bottomrule
  \end{tabular}
\end{table}

The missing-value rate (GT\,=\,0.05, exactly at the floor) incurs only 2.7\% MRE,
confirming that RU estimates well-represented quality dimensions accurately.
The aggregate MRE of 0.401 is therefore driven entirely by the three
floor-dominated dimensions, not by any real failure of the sampler on D1.

Table~\ref{tab:indicator} provides a per-indicator breakdown confirming that
\texttt{dag}'s error on D2 and D3 concentrates entirely in the outlier column
(MRE\,=\,0.682 and 0.846 respectively) rather than the missing, duplicate, or
FD columns, identifying IQR threshold mismatch as the sole driver of failure.

\begin{table}[!htbp]
  \small
  \centering
  \caption{Per-indicator mean MRE on D2 (NYC 311) and D3 (NYPD arrests),
           averaged across all budgets and seeds.
           Only Miss (missing) and Out (outlier) are shown;
           Dup and FD are identically zero across all methods in both datasets.
           Bold = lowest MRE per column.}
  \label{tab:indicator}
  \begin{tabular}{l rr rr}
    \toprule
    & \multicolumn{2}{c}{D2 (NYC 311)} & \multicolumn{2}{c}{D3 (NYPD)} \\
    \cmidrule(lr){2-3}\cmidrule(lr){4-5}
    Method & Miss & Out & Miss & Out \\
    \midrule
    \texttt{RU}         & \textbf{.002} & \textbf{.015} & \textbf{.000} & .074 \\
    \texttt{cluster}    & .008 & .020 & .000 & \textbf{.035} \\
    \texttt{yamane}     & .007 & .374 & .003 & .215 \\
    \texttt{dag}        & .099 & .682 & .007 & .846 \\
    \texttt{MH}         & .092 & .674 & .007 & .835 \\
    \texttt{geom.}      & \textbf{.002} & \textbf{.015} & \textbf{.000} & .074 \\
    \texttt{strat-col}  & 1.000 & .682 & .023 & .957 \\
    \texttt{strat-q}    & \textbf{.002} & \textbf{.015} & \textbf{.000} & .074 \\
    \texttt{importance} & .271 & .682 & .007 & .846 \\
    \bottomrule
  \end{tabular}
\end{table}

\subsection{E3 --- Budget--Accuracy Trade-off}
\label{sec:e3}

Table~\ref{tab:e3} shows mean relative error of \texttt{random\_uniform} and
\texttt{dag} at each sampling budget on D2, D3, and D4.

\begin{table}[!htbp]
  \small
  \centering
  \caption{E3 --- Mean relative error vs.\ budget (3 seeds/cell). \texttt{dag}
  is budget-invariant at 19--27\%. Bold = lowest \texttt{RU} error per dataset
  (D2 at 5\%, C2; D3 and D4 at 50\%). The bolded D2/\texttt{RU} value at $b$=5\%
  is 0.0049 (0.49\%, C2), shown rounded to .005.}
  \label{tab:e3}
  \begin{tabular}{ll rrrrr}
    \toprule
    Dataset & Method & $b$=5\% & $b$=10\% & $b$=20\% & $b$=30\% & $b$=50\% \\
    \midrule
    \multirow{2}{*}{D2 (NYC 311)}
      & \texttt{RU}  & \textbf{.005} & .007 & .004 & .003 & .003 \\
      & \texttt{dag} & .195          & .196 & .196 & .195 & .194 \\
    \midrule
    \multirow{2}{*}{D3 (NYPD)}
      & \texttt{RU}  & .037          & .024 & .020 & .010 & \textbf{.001} \\
      & \texttt{dag} & .213          & .213 & .213 & .213 & .213 \\
    \midrule
    \multirow{2}{*}{D4 (Adult)}
      & \texttt{RU}  & .027          & .025 & .021 & .018 & \textbf{.011} \\
      & \texttt{dag} & .273          & .274 & .260 & .262 & .264 \\
    \bottomrule
  \end{tabular}%
\end{table}

\begin{figure}[!htbp]
  \centering
  \includegraphics[width=\linewidth]{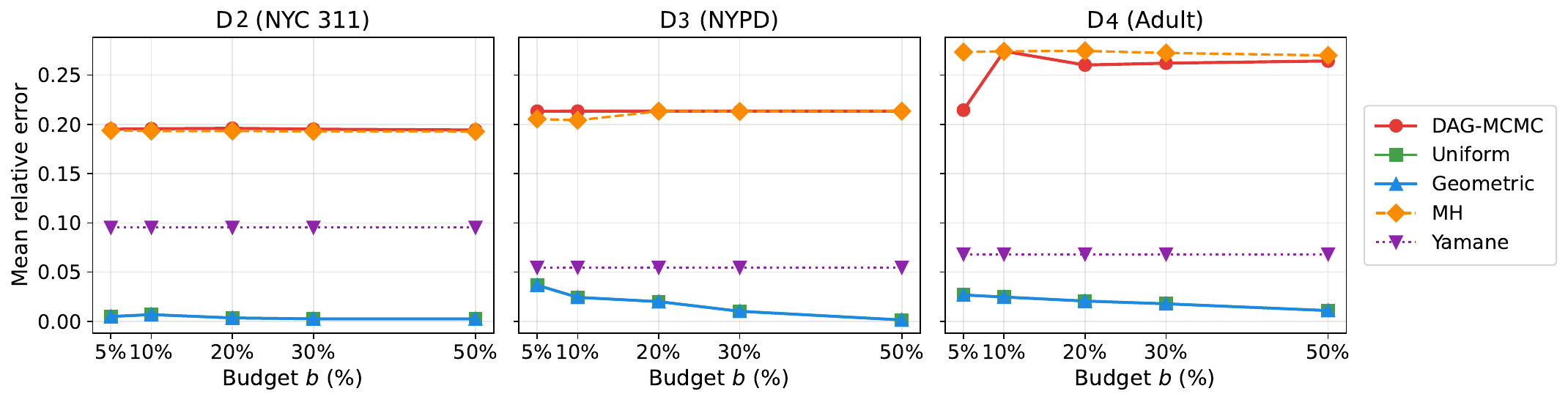}
  \caption{E3 --- Relative error vs.\ sampling budget $b$.
           D2=NYC~311; D3=NYPD arrests; D4=UCI Adult census.
           \texttt{dag} is budget-invariant at $\approx$19--27\%;
           \texttt{RU} converges to sub-1\% at $b$=5\% on D2.}
  \label{fig:e3}
\end{figure}

\textbf{Finding (C2).}
On D2 (NYC 311, 500K rows), \texttt{random\_uniform} reaches \textbf{0.49\%
mean relative error at a 5\% sampling budget}, far below the 1\% threshold with
only 25K rows inspected.
\texttt{strat-quality} matches this on D2 (0.49\% at $b$=5\%); \texttt{cluster}
reaches 1.42\% at $b$=5\% and converges to 0.21\% at $b$=50\%
(Table~\ref{tab:e3} lists \texttt{RU} and \texttt{dag}; per-budget values for all
nine methods are in the released results).
\texttt{dag} is insensitive to budget on all three real datasets (19--27\%
regardless of $b$), confirming that its proxy mismatch is structural
(Figure~\ref{fig:e3}).

\subsection{E4 --- Robustness under Error Injection}
\label{sec:e4}

We inject four error types (missing, duplicate, outlier, inconsistency) at rates
1\%--30\% into D1 and measure mean relative error (3 seeds per cell, $b$=0.50).
Table~\ref{tab:e4} reports means aggregated over injection rates 1\%--30\%.

\begin{table}[!htbp]
  \small
  \centering
  \caption{E4 --- Mean relative error under error injection (rates 1--30\%,
           15 obs/cell). \texttt{random\_uniform} (RU) is 5.6$\times$ more robust
           than \texttt{dag}.}
  \label{tab:e4}
  \begin{tabular}{lrr}
    \toprule
    Error type     & \texttt{dag} & \texttt{RU} \\
    \midrule
    Missing        & .438        & .051 \\
    Duplicate      & .342        & .112 \\
    Outlier        & .349        & .046 \\
    Inconsistency  & .332        & .050 \\
    \midrule
    \textbf{Mean}  & \textbf{.365} & \textbf{.065} \\
    \bottomrule
  \end{tabular}
\end{table}

\begin{figure}[!htbp]
  \centering
  \subfloat[\label{fig:e4}]{%
    \includegraphics[height=80pt]{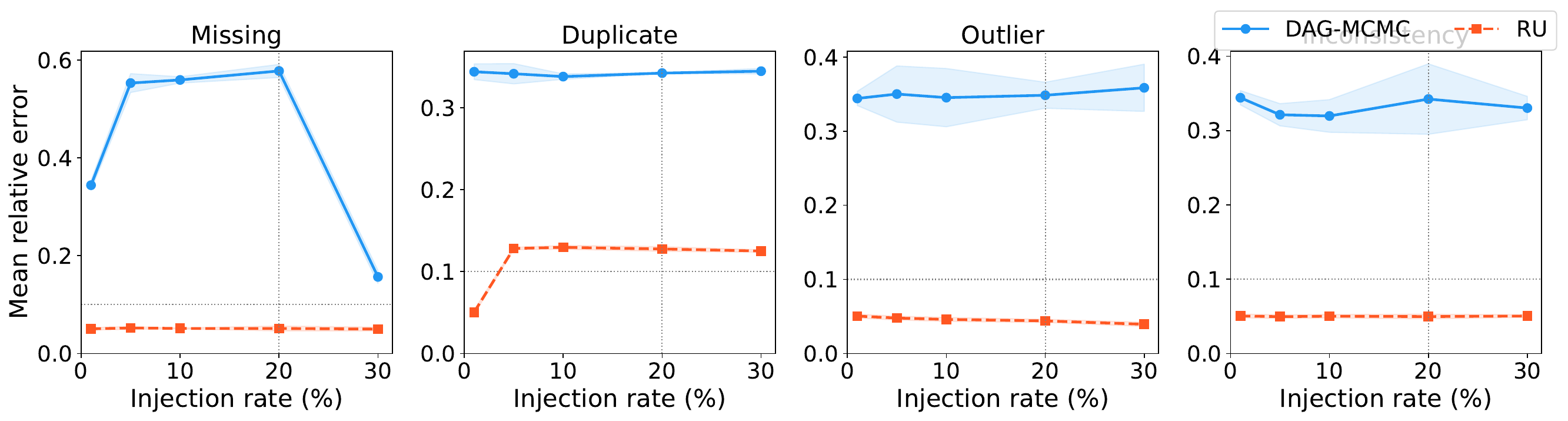}}\\[4pt]
  \subfloat[\label{fig:e5}]{%
    \includegraphics[height=80pt]{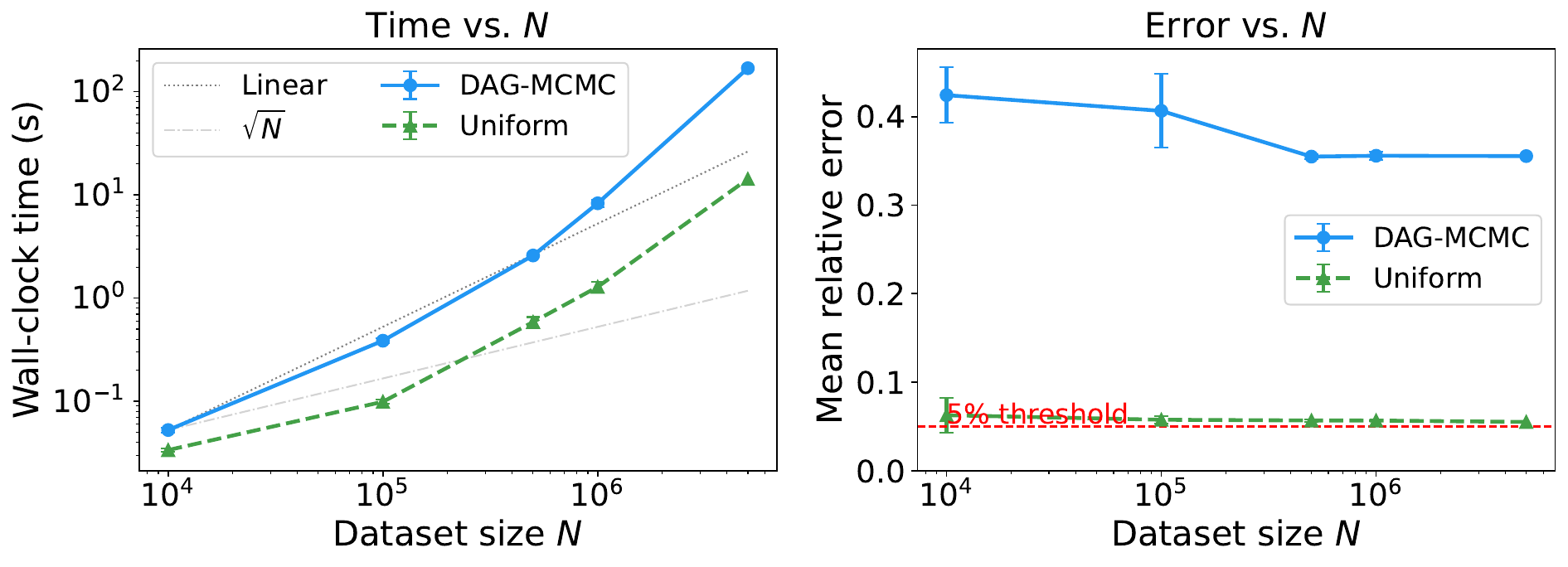}}\hfill
  \subfloat[\label{fig:e2}]{%
    \includegraphics[height=80pt]{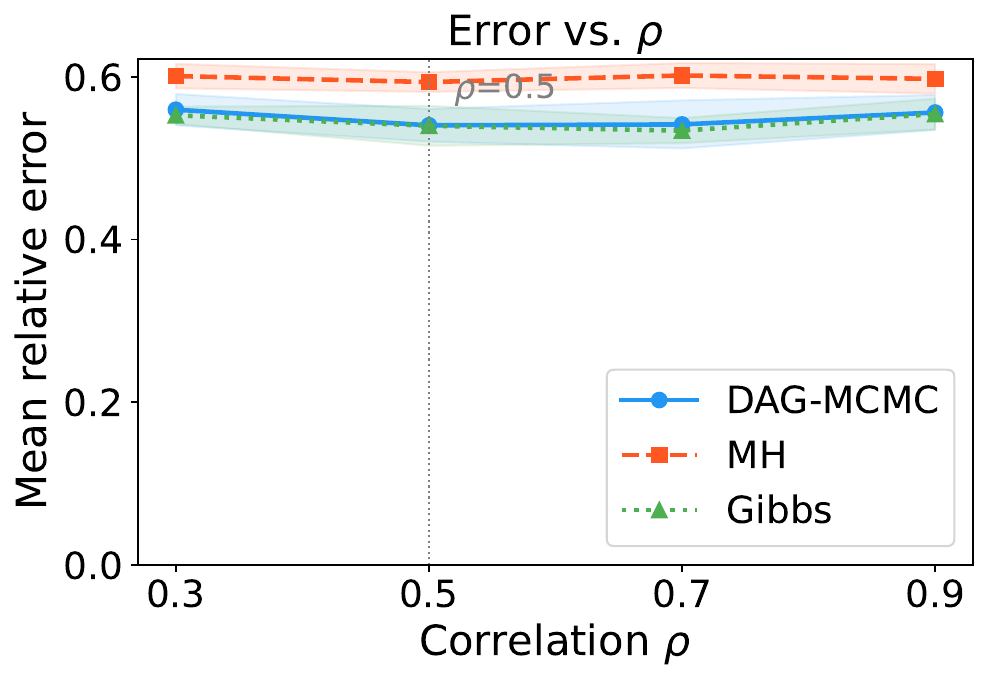}}
  \caption{(a) E4 --- Robustness: MRE vs.\ error injection rate (\%, D1 synthetic,
  dag vs.\ RU, 3 seeds). (b) E5 --- Scalability: wall-clock time (s) vs.\ $N$
  rows (D5 SDV synthetic, dag vs.\ RU, 1--3 seeds; see footnote). (c) E2 ---
  Correlation ablation: MRE vs.\ Pearson $\rho$ (D1 synthetic, dag vs.\ MH
  vs.\ Gibbs, 5 seeds per $\rho$). Lower MRE is better.}
  \label{fig:e4e5e2}
\end{figure}

\textbf{Finding (C5).}
\texttt{random\_uniform} (RU) maintains a mean error of 0.065 across all error
types and injection rates (range: 0.04--0.13), while \texttt{dag} reaches 0.365
(range: 0.33--0.44), a 5.6$\times$ gap (Figure~\ref{fig:e4}).

\subsection{E5 --- Scalability }
\label{sec:e5}

We profile D5 at five scales $N \in \{10\text{K}, 100\text{K}, 500\text{K},
1\text{M}, 5\text{M}\}$%
\footnote{The $N$=5M condition uses 1 seed due to wall-time constraints
($\approx$170\,s per \texttt{dag} run); the point is consistent with the trend
at $N$=1M.}
and fit a power law $T = aN^{\alpha}$ on log-log axes.
Table~\ref{tab:e5} reports wall-clock time and mean relative error.

\begin{table}[!htbp]
 \small
  \centering
  \caption{E5 --- Wall-clock time (s) and error vs.\ scale on D5 ($b$=0.50).
           Power-law exponents: \texttt{RU} $\alpha$=0.964; \texttt{dag} $\alpha$=1.272.}
  \label{tab:e5}
  \begin{tabular}{l rrrrr}
    \toprule
    Method & 10K & 100K & 500K & 1M & 5M \\
    \midrule
    \multicolumn{6}{l}{\textit{Wall-clock time (seconds)}} \\
    \texttt{dag}        & .050  & .390  & 2.59   & 8.29   & 169.0 \\
    \texttt{RU}         & .030  & .100  & .590   & 1.28   & 14.3  \\
    \texttt{exhaustive} & .030  & .150  & .860   & 1.97   & 18.9  \\
    \midrule
    \multicolumn{6}{l}{\textit{Mean relative error}} \\
    \texttt{dag}        & .424 & .407 & .355  & .356  & .355 \\
    \texttt{RU}         & .063 & .058 & .057  & .057  & .055 \\
    \texttt{exhaustive} & .000 & .000 & .000  & .000  & .000 \\
    \bottomrule
  \end{tabular}
\end{table}

\textbf{Finding (C4).}
\texttt{random\_uniform} scales with a near-linear cost exponent ($\alpha$=0.964)
and achieves stable error around 5.5--6.3\% across all five scales
(Figure~\ref{fig:e5}).
At $N$=5M, \texttt{RU} completes in 14.3\,s---\textbf{faster than exhaustive
profiling} (18.9\,s) while incurring only 5.5\% error.
\texttt{dag} scales super-linearly ($\alpha$=1.272): at $N$=5M it requires
169.0\,s, 12$\times$ slower than \texttt{RU} and 6$\times$ less accurate.

\subsection{E2 --- Correlation Ablation }
\label{sec:e2}

We assess whether attribute correlation strength modulates MCMC sampling quality
on D1 by varying $\rho \in \{0.3, 0.5, 0.7, 0.9\}$ (5 seeds per $\rho$,
$b$=0.50).
Table~\ref{tab:e2} reports mean relative error per method and $\rho$.

\begin{table}[!htbp]
  \small
  \centering
  \caption{E2 --- Mean relative error vs.\ correlation $\rho$ on D1 (5 seeds,
  $b$=0.50). DAG shows a consistent directional advantage (lower MRE than MH at
  all $\rho$ levels) but the effect does not meet the Bonferroni-corrected
  significance threshold ($\alpha/4 = 0.0125$); the result is reported as
  open/inconclusive (C6).}
  \label{tab:e2}
  \begin{tabular}{l rrr r}
    \toprule
    $\rho$ & \texttt{dag} & \texttt{gibbs} & \texttt{MH}
           & $p$(\texttt{dag}$<$\texttt{MH}), $n$=5 \\
    \midrule
    .300  & .560 & .553 & .601 & .031 \\
    .500  & .540 & .540 & .594 & .031 \\
    .700  & .542 & .534 & .602 & .031 \\
    .900  & .556 & .554 & .598 & .031 \\
    \midrule
    Mean & .550 & .545 & .599 & -- \\
    \bottomrule
  \end{tabular}
  \vspace{2pt}
  \noindent\footnotesize
  $p$=0.03125 ($= 1/2^5$) is the minimum achievable one-sided Wilcoxon $p$-value
  with $n$=5. Direction is consistent: \texttt{dag} beats \texttt{MH}
  in all 20 seed-$\rho$ pairs; \texttt{gibbs} beats \texttt{MH} in 21/25 pairs.
  Tests are underpowered per stratum.
\end{table}

\textbf{Finding (C6) --- open question.}
Both \texttt{dag} (0.550) and \texttt{gibbs} (0.545) are directionally better
than \texttt{MH} (0.599) at all 20 seed-$\rho$ pairs each (8--9\% relative reduction),
but per-$\rho$ tests are underpowered: with $n$=5 seeds, the minimum achievable
one-sided Wilcoxon $p$-value is $1/2^5 = 0.03125$, which does not meet the
Bonferroni-corrected threshold ($\alpha/4 = 0.0125$).
We treat this as inconclusive directional evidence requiring $n \geq 20$ seeds
(Figure~\ref{fig:e2}).

\subsection{E6 --- XXL Real-Dataset Scalability}
\label{sec:e6}

We profile two XXL real datasets: D6-A (7{,}461{,}195 rows) and D6-B
(5{,}980{,}721 rows).
Ground truth is computed exhaustively once and cached; 10 seeds per condition.
Table~\ref{tab:e6} reports mean relative error for both datasets.

\begin{table}[!htbp]
 \small
  \centering
  \caption{E6 --- Mean relative error (\%) on two XXL real datasets (10 seeds/cell),
           with mean wall-clock time (s) at each budget.
           \texttt{dag} is consistently high and far from RU regardless of budget
           or dataset. \texttt{geometric} omitted: at these budgets it degenerates
           to RU (identical values).}
  \label{tab:e6}
  \begin{tabular}{llrrrrr}
    \toprule
    Dataset & Method & $b$=1\% & $b$=2\% & $b$=5\% & $b$=10\% & $b$=20\% \\
    \midrule
    \multirow{2}{*}{D6-A (7.4M)}
      & \texttt{RU}        & \textbf{1.27} & \textbf{.560} & \textbf{2.03} & \textbf{1.10} & \textbf{.710} \\
      & \texttt{dag}       & 18.3          & 23.9          & 24.3          & 25.5          & 25.4 \\
    \midrule
    \multirow{2}{*}{D6-B (6.0M)}
      & \texttt{RU}        & \textbf{.420} & \textbf{.350} & \textbf{.200} & \textbf{.110} & \textbf{.060} \\
      & \texttt{dag}       & 17.4          & 17.3          & 17.2          & 17.4          & 17.6 \\
    \midrule
    \multicolumn{7}{l}{\textit{Wall-clock time (s, mean over 10 seeds)}} \\
    \multirow{2}{*}{D6-A (7.4M)}
      & \texttt{RU}  & .240 & .380 & .800 & 1.46 & 2.72 \\
      & \texttt{dag} & 8.92 & 12.24 & 22.19 & 40.63 & 79.38 \\
    \multirow{2}{*}{D6-B (6.0M)}
      & \texttt{RU}  & .160 & .260 & .610 & 1.24 & 2.10 \\
      & \texttt{dag} & 14.57 & 17.90 & 27.58 & 42.96 & 74.77 \\
    \bottomrule
  \end{tabular}%
\end{table}

\textbf{Finding (C4 --- XXL extension).}
On both XXL datasets, \texttt{random\_uniform} achieves well under 5\% mean
relative error at 5\% budget (2.03\% on D6-A; 0.20\% on D6-B).
At 5\% budget, \texttt{RU} completes in 0.80\,s (D6-A) and 0.60\,s (D6-B) while
\texttt{dag} requires 22.19\,s and 28.13\,s respectively, yielding timing gaps of
28$\times$ (D6-A) and \textbf{47$\times$} (D6-B)---both larger
than the 12$\times$ at $N=5\text{M}$ synthetic, confirming super-linear DAG
scaling on real data.

\subsection{E8 --- IoT Sensor Benchmark}
\label{sec:e8}

We profile two IoT datasets: D7-synth (510K synthetic rows, injected defects:
5.4\% missing, 2.0\% duplicates, 16.0\% outliers, 2.9\% FD inconsistencies)
and D7-real (2.3M rows from the Intel Berkeley Research Lab~\cite{tolle2005macroscope};
ground truth computed exhaustively: 0.4\% missing, 0.0\% duplicates,
29.6\% outliers, 0.0\% FD inconsistencies).
Outliers arise from genuine sensor malfunctions (temperature spikes, voltage
drops)---a numeric-dominant quality profile absent in all prior datasets.
Table~\ref{tab:e8} reports mean relative error (MRE, \%) at 5\% and 50\%
budgets over 3 seeds.

\begin{table}[!htbp]
 \small
  \centering
  \caption{E8 --- Mean relative error (\%) on IoT datasets, 3 seeds/cell.
           Bold = lowest MRE per row.
           \texttt{cluster} matches \texttt{RU}/\texttt{geom.}\ on D7-synth;
           \texttt{strat-q} matches \texttt{RU} on D7-real (uniform outlier distribution).
           \texttt{dag}, \texttt{MH}, \texttt{strat-col}, \texttt{importance} fail on D7-real
           via IQR estimation collapse (see Section~\ref{sec:discussion}).}
  \label{tab:e8}
  \begin{tabular}{lrrrr}
    \toprule
    Method & \multicolumn{2}{c}{D7-synth (510K)} & \multicolumn{2}{c}{D7-real (2.3M)} \\
    \cmidrule(lr){2-3}\cmidrule(lr){4-5}
           & $b$=5\% & $b$=50\% & $b$=5\% & $b$=50\% \\
    \midrule
    \texttt{RU}         & \textbf{23.5} & \textbf{12.2} & \textbf{.230} & \textbf{.180} \\
    \texttt{geom.}      & \textbf{23.5} & \textbf{12.2} & \textbf{.230} & \textbf{.180} \\
    \texttt{yamane}     & 28.5 & 28.5 & 1.33 & 1.33 \\
    \texttt{cluster}    & \textbf{23.5} & \textbf{11.9} & 2.36 & .470 \\
    \midrule
    \texttt{dag}        & 72.6 & 59.7 & 26.9 & 26.9 \\
    \texttt{MH}         & 71.9 & 57.6 & 26.8 & 26.9 \\
    \texttt{strat-q}    & 53.0 & 45.0 & \textbf{.230} & \textbf{.180} \\
    \texttt{strat-col}  & 53.0 & 45.0 & 22.8 & 22.8 \\
    \texttt{importance} & 48.5 & 35.7 & 19.0 & 26.9 \\
    \bottomrule
  \end{tabular}%
\end{table}

\textbf{Finding (C7 --- IoT sensor benchmark).}
On D7-real, \texttt{random\_uniform}, \texttt{geom.}, and \texttt{strat-q} all
achieve 0.23\% MRE at 5\% budget---below the 1\% threshold of C2 on a dataset
4.6$\times$ larger than NYC~311 ($14.3\times$ speedup for RU).
\texttt{cluster} starts at 2.36\% MRE at 5\% budget but converges to 0.47\% at 50\%.
\texttt{dag} reports 26.9\% MRE across \emph{all} budgets including 100\%, while running
11$\times$ \emph{slower} than exhaustive; \texttt{strat-col} (22.8\%) and
\texttt{importance} (19.0\%) fail via the same IQR estimation collapse.
On D7-synth, RU, geometric, and cluster (all $\approx$12--23\% at 5--50\%)
outperform DAG (59.7\%), MH (57.6\%), and the IQR-proxy-stratified methods
(strat-q: 45--53\%) by a factor of 2--4.
With $n$=3 seeds, the minimum achievable one-sided Wilcoxon $p$-value is
$1/2^3 = 0.125$; E8 results are therefore reported descriptively rather than
with formal significance, consistent with the per-stratum treatment in E2.
The point estimates (26.9\% vs.\ 0.23\% MRE, a ratio of ${\approx}117{\times}$)
are large enough to be practically unambiguous, but formal statistical
confirmation requires $n \geq 7$ seeds.

The DAG failure on D7-real is mechanistically different from the categorical proxy
mismatch on administrative data (Section~\ref{sec:discussion}): it is an
\emph{estimation collapse}.
DAG's non-uniform sample over-represents rows near IQR boundaries;
in the resulting sample, the empirical Q1/Q3 shift outward, widening the
sample-level IQR so that almost no sampled row is flagged as an outlier.
The estimated outlier rate collapses to $\approx 3\times10^{-6}$ against a
true rate of 29.6\%, contributing a per-indicator relative error of $\approx 1.0$
that dominates the MRE.
Random uniform sampling, in contrast, draws a representative cross-section whose
empirical IQR matches the population, yielding an estimated outlier rate of
29.3\% (true: 29.6\%) at every tested budget.

\paragraph{Practical recommendation for DC-AI pipelines.}
For IoT or sensor streaming data where quality defects are predominantly
\emph{numeric and uniformly distributed} across devices, random uniform sampling
at a 5--10\% budget is the strongly preferred choice.
Markov chain methods add substantial overhead (11--13$\times$ slower) without
accuracy benefit; their advantage---if any---requires data where defects are
spatially concentrated and structurally encoded, conditions not met in real
sensor deployments.

\section{Discussion}
\label{sec:discussion}

\paragraph{Why does guided sampling fail on real data?}
Five failure modes explain the consistent underperformance of proxy-guided methods;
two controlled ablation experiments (E7 a and b) confirm the attribution.

\emph{IQR proxy mismatch.}
The IQR-based error proxy detects numeric outliers---extreme latitude/longitude
values in NYC~311 (D2/D3) or capital\_gain/loss spikes in UCI Adult (D4).
Quality defects in all three real datasets, however, concentrate in categorical
and string-typed columns.
This mismatch is structural: any IQR-style proxy is blind to categorical data
quality by construction~\cite{hellerstein2008quantitative}.

\emph{Horvitz-Thompson weighting instability.}
When row scores cluster near zero (string-heavy datasets), importance weights
$w_i \propto 1/\text{score}_i$ diverge toward $10^6$.
Even after clipping, over-weighted rows dominate the estimate and bias all
indicators toward zero.

\emph{Proxy ablation (E7a).}
\begin{sloppypar}
A controlled ablation (\texttt{dag\_uniform}) retains DAG's graph proposal but
replaces IQR-based HT weights with uniform weights.
On D2 at $b$=50\%, \texttt{dag\_uniform} achieves 3.8\% MRE vs.\ DAG's
persistent 19.5\%---confirming the IQR proxy introduces a structural bias.
We note one non-monotonic exception: at $b=5\%$ on D2,
\texttt{dag\_uniform} achieves 30.5\% MRE---worse than full \texttt{dag}
($19.6\%$)---indicating that at very low budget, the IQR weights partially
compensate for the proposal bias introduced by the DAG structure.
The dominance of the weighting mechanism over the proposal bias is therefore
budget-dependent: it is negligible at $b \leq 5\%$ and pronounced at
$b \geq 10\%$ (\texttt{dag\textsubscript{uniform}} $= 3.8\%$ vs.\
\texttt{dag} $= 19.5\%$ at $b=50\%$).
This motivates a refined root-cause framing: \emph{the IQR weighting step is
the dominant failure source at operational budgets ($b \geq 10\%$); at very
low budget, the DAG proposal and weighting interact and partially cancel.}
Table~\ref{tab:e7} reports the full ablation.
\end{sloppypar}

\emph{Hyperparameter robustness (E7b).}
A sweep over MCMC batch size $B \in \{100, 200, 500, 1000\}$ on D2 confirms
that DAG's MRE varies by less than 0.001 across all four values (19.48--19.57\%);
the reported $B$=500 default is not a cherry-picked optimum.
Wall-clock time decreases with larger $B$ (2.56\,s at $B$=100 to 0.90\,s at
$B$=1000) but accuracy is flat---the proxy mismatch dominates regardless of step
granularity (detailed results available in released code).

\begin{table}[!htbp]
\scriptsize
\centering
\caption{E7a --- Proxy ablation: \texttt{dag\_uniform} retains DAG's graph
proposal but replaces IQR-based HT weights with uniform weights.
Columns: \texttt{dag} = full DAG-guided MCMC (IQR proxy + graph);
\texttt{dag\_unif.} = graph proposal only (uniform weights); RU = random uniform.
Means over 3 seeds. Bold = lowest MRE per row.}
\label{tab:e7}
\begin{tabular}{llrrr}
\toprule
\textbf{Dataset} & \textbf{$b$ (\%)} & \textbf{\texttt{dag}} & \textbf{\texttt{dag\_unif.}} & \textbf{RU} \\
\midrule
D2 (NYC 311) & 5  & .196 & .305          & \textbf{.005} \\
D2 (NYC 311) & 50 & .195 & .038          & \textbf{.003} \\
D3 (NYPD)    & 5  & .213 & \textbf{.024} & .037 \\
D3 (NYPD)    & 50 & .213 & .009          & \textbf{.001} \\
\bottomrule
\end{tabular}
\end{table}

\emph{IQR estimation collapse (IoT data).}
On D7-real (29.6\% true outlier rate), the failure mechanism is distinct from
categorical proxy mismatch.
DAG's non-uniform sample concentrates extreme-value rows, shifting the sample's
Q1/Q3 outward and widening the IQR so that virtually no sampled row is classified
as an outlier.
The estimated outlier rate collapses to $\approx 3\times10^{-6}$, yielding
$\approx 1.0$ per-indicator relative error that persists even at 100\% budget---
confirming this is an estimator bias, not a sampling coverage issue.

\emph{Proxy-guided non-representativeness generalises beyond MCMC.}
The new stratified and importance methods confirm that the root cause is not
MCMC-specific: any sampler that concentrates draws on high-IQR-proxy rows
shifts the sample IQR outward and biases all downstream estimates.
\texttt{strat-col} (Neyman allocation by column type + IQR proxy) and
\texttt{importance} (weighted reservoir sampling $\propto$ proxy score) both
achieve primary mean MRE 0.31--0.35---comparable to dag/MH---because they
amplify the same proxy mismatch.
By contrast, \texttt{cluster} (random consecutive blocks, no proxy) matches RU
(primary mean 0.110 vs.\ 0.111), establishing that \emph{representativeness},
not domain knowledge, is the decisive property for multi-indicator quality profiling.

\emph{Super-linear computational cost.}
At each batch step, DAG resamples the entire candidate pool, making total cost
super-linear in $N$ ($O(N^{1.272})$ empirically).
At $N=5\text{M}$ rows this is 169\,s against RU's 14.3\,s (12$\times$ slowdown
with 6$\times$ worse accuracy).
On XXL real data the gap widens further: 28$\times$ on D6-A and 47$\times$
on D6-B.

\paragraph{Implications for DC-AI practitioners.}
Data quality profiling occupies three positions in a DC-AI pipeline:
\emph{ingestion-time} (profile each arriving batch), \emph{pre-training}
(verify the full training set before a model fit), and \emph{drift monitoring}
(compare the current profile against a stored reference baseline).
Our results point to the same configuration in all three settings:
use \texttt{random\_uniform} or \texttt{cluster} at a 5--10\% budget.

At \emph{ingestion-time}, a 5\% budget over a 500K-row daily update scans
only 25K rows, achieving $<$1\% MRE on all four quality indicators with
latency negligible against typical ingestion windows.
At \emph{pre-training}, the same budget over a 5M-row training corpus requires
scanning 250K rows, completing in under 15 seconds on commodity hardware
(vs.\ 169 seconds for a full scan; E5)---freeing the remainder of the training
budget for model iteration rather than data scanning.
For \emph{drift monitoring}, the profile estimate is precise enough ($<$1\% MRE
on real data) to detect meaningful shifts in missing-value rates, duplicate
fractions, and outlier densities between pipeline runs.

Crucially, neither \texttt{random\_uniform} nor \texttt{cluster} requires any
schema metadata, dependency graph, or domain-specific tuning.
This schema-free property is essential in DC-AI systems where data sources are
heterogeneous and schemas evolve: the same profiler applies without
reconfiguration to administrative tables, IoT sensor streams, and census
microdata.
DAG-guided methods, by contrast, require constructing an attribute correlation
graph---a non-trivial data engineering effort that must be repeated when the
schema changes, and which our results demonstrate adds no accuracy benefit on
any real dataset tested.
Under error injection RU is also the most robust choice ($<$13\% MRE at
injection rates up to 30\%; E4), making it the safe default under both clean
and corrupted conditions.

The data-type-specific lesson is as follows: on \emph{administrative} data with
categorical defects the IQR proxy is structurally mismatched; on \emph{numeric
IoT} data with high outlier rates the IQR estimator collapses; in both cases
any proxy-guided method adds overhead without benefit.
The decisive property for multi-indicator quality profiling is
\emph{representativeness}---drawing a sample whose empirical distribution
mirrors the population---not domain knowledge encoded in a proxy or graph.
Whether genuinely structured relational or knowledge-graph data changes this
conclusion remains an open question (E2).

\paragraph{Limitations.}
Real-world data evaluation covers administrative, census, and IoT sensor tabular data;
time-series (beyond single-stream IoT), knowledge-graph,  multi-relational and multimodal 
settings are out of scope.
All experiments use 3--10 seeds~\cite{bouthillier2021accounting}. Our open finding (C6) would require $n \geq 20$ seeds for confirmation.
\section{Related Work}
\label{sec:related}

\paragraph{Data profiling and quality assessment.}
Data profiling encompasses automated discovery of metadata, structural properties,
and quality indicators from datasets.
Abedjan et al.\ provide a comprehensive taxonomy and benchmark of profiling
systems~\cite{abedjan2015profiling,abedjan2018data}.
Schelter et al.\ present an automated large-scale data quality verification system
integrating declarative constraint checking into ML pipelines~\cite{schelter2019}.
Grafberger et al.\ extend this to runtime distribution debugging across pipeline
stages, monitoring data slices between training and serving~\cite{grafberger2022dist};
our work targets the complementary pre-deployment question of how accurately
quality profiles can be estimated from a small fraction of the data.
Naumann and Herschel lay the theoretical foundations of data quality
dimensions~\cite{naumann2014data}.
Closely related to our relational-consistency indicator, Berti-\'Equille et
al.~\cite{func_dep} study the discovery of genuine functional dependencies from
relational data with missing values, a prerequisite for the FD-violation defects
our profile measures.
On the repair side, Rekatsinas et al.~\cite{rekatsinas2017holoclean} perform
holistic data repair with probabilistic inference.
Whang et al.\ survey data collection and quality challenges in ML pipelines from
a DC-AI perspective, identifying heterogeneous quality failures across column types
as primary obstacles to reliable model training~\cite{whang2023data}---the
same column-type heterogeneity that explains the IQR proxy mismatch we identify
in Section~\ref{sec:discussion}.
Berti-Equille examines the interaction between data quality and downstream
analytical costs~\cite{bertie_dq_aware}---the same cost--quality trade-off
motivating our progressive profiling approach.
A critical limitation shared by all these systems is that they operate
\emph{exhaustively}: the full dataset must be processed before any quality
indicator is reported.
Our work addresses this gap by studying how well quality profiles can be estimated
from progressively drawn samples.

\paragraph{Progressive and approximate query processing.}
The approximate query processing (AQP) literature addresses the cost of exact
query evaluation.
Hellerstein et al.\ pioneer online aggregation~\cite{hellerstein1997online};
Haas and Hellerstein extend this to joins~\cite{haas1999ripple}.
BlinkDB~\cite{agarwal2013blinkdb} and Verdict~\cite{park2017verdict} build
pre-computed stratified samples for ad-hoc analytical queries.
Most recently, Zhu proposes B-AQP, which uses block-level (cluster) sampling
for AQP and achieves a 41\% error reduction over uniform sampling on aggregate
queries~\cite{zhu2025block}; their finding that block draws are competitive with
uniform sampling on aggregate estimation aligns with our result that our
\texttt{cluster} sampler matches random uniform on multi-indicator quality profiling.
However, AQP systems target aggregate queries over a single statistic; data
quality profiling involves simultaneously estimating multiple heterogeneous
indicators, leaving open the question of which sampling strategy best suits
this multi-indicator setting---a question we answer empirically.

\paragraph{Sampling for data quality.}
\begin{sloppypar}
Cormode et al.\ survey synopses for massive datasets~\cite{cormode2012synopses}.
Vitter's reservoir sampling~\cite{vitter1985random} and weighted extensions of
Efraimidis and Spirakis~\cite{efraimidis2006weighted} provide the algorithmic
foundation for importance-weighted sampling underlying our MCMC-guided methods;
our experiments ask whether these importance weights actually help for
multi-indicator profiling, and find they do not on real administrative data.
\end{sloppypar}

\paragraph{Stratified and adaptive sampling.}
Stratified random sampling~\cite{neyman1934two} allocates sample sizes to strata
proportionally to their variance.
Adaptive sampling~\cite{thompson1990adaptive} dynamically adjusts stratum
boundaries.
We benchmark two stratified variants: \texttt{strat-col} partitions rows by column
type and IQR proxy, and \texttt{strat-quality} partitions by IQR proxy quantiles,
both using Neyman allocation.
Contrary to expectation, both inherit the IQR proxy mismatch and match or exceed
dag/MH error on real data; only \texttt{strat-quality} reaches RU-level accuracy on
D2 and D7-real where quality defects happen to correlate with the proxy.

\paragraph{Foundational progressive sampling.}
Provost et al.~\cite{provost1999efficient} established that geometric growth
schedules are efficient for learning curves.
John and Langley~\cite{john1996static} showed that dynamic sampling outperforms
static budgets.
The present work applies the same principle to quality profiling metrics and
evaluates it across nine strategies on real administrative and IoT data at scale.

\paragraph{MCMC row-sampling for data quality profiling.}
MCMC methods for \emph{row-level} selection in data quality profiling are largely
unexplored in the literature.
Existing work, including the comprehensive survey and benchmark of Abedjan
et al.~\cite{abedjan2018data}, focuses on \emph{column-level} dependency analysis
and automated error detection, not on row-level sampling strategies for profiling
estimation.
Our DAG-guided sampler is therefore an original contribution that bridges
attribute-level dependency modeling---inspired by the column-correlation perspective
of Abedjan et al.---and row-level progressive sampling: it uses the attribute DAG
to shape a Metropolis--Hastings proposal distribution over rows, weighted by an
IQR-based error proxy.
Empirically, this bridge does not improve profiling accuracy over blind sampling
on real tabular data, a finding that holds across administrative, census, and IoT
data types and constitutes a reproducible negative result for structure-aware row
samplers.

\paragraph{Our contribution in context.}
To our knowledge, no prior work systematically compares progressive sampling
strategies for \emph{multi-indicator} data quality profiling at scale.
The data-centric AI movement~\cite{Mazumder2022,zha2024dcai,bertie_explai_ai},
part of a broader round trip between machine learning and data
management~\cite{bertie_ml_dm}, has renewed interest in profiling pipelines and
raised the expectation that structure-aware, dependency-guided samplers should
outperform blind draws.
Our work overturns this assumption: the IQR-based proxy that DAG-guided methods
rely on is blind to categorical quality defects, making random uniform sampling
faster and more accurate on all real datasets tested---a reproducible negative
result with direct implications for production DC-AI data monitoring pipelines.

\section{Conclusion}
\label{sec:conclusion}

Data-centric AI places data quality at the centre of the ML lifecycle, yet the
question of \emph{how} to measure that quality efficiently at scale has received
little systematic attention.
We have shown that \emph{representativeness}, not domain knowledge or quality
proxies, is the decisive property for multi-indicator data quality profiling.
Across six real datasets---three administrative (NYC~311 and NYPD arrests, 500K rows each; UCI Adult census, 49K rows), two XXL tabular (D6-A: 7.4M rows, D6-B: 6.0M rows), and one IoT sensor stream (D7-real: 2.3M rows)---plus synthetic tabular data scaled to 5M rows, blind representative samplers (random uniform, cluster) dominate all
nine benchmarked strategies on accuracy, cost, and robustness.
DAG-guided MCMC is 11--49$\times$ less accurate on real datasets and 12$\times$
slower at $N=5$M; stratified and importance-weighted methods that rely on
IQR-based proxies inherit the same structural failure (strat-col and importance: MRE 0.31--0.35; strat-quality intermediate at 0.20).
Cluster sampling---which draws random consecutive blocks without any proxy---matches
random uniform (primary mean MRE 0.110 vs.\ 0.111) with $O(N)$ cost and no tuning.

\textbf{Practical recommendation for DC-AI pipelines.}
Use \texttt{random\_uniform} or \texttt{cluster} at a 5--10\% sampling budget for
routine data quality monitoring.
\texttt{random\_uniform} is also the most robust baseline under error injection
across all four error types (mean error 0.065 vs.\ 0.365 for DAG) at injection
rates up to 30\% (E4), making it the default choice under both normal and
error-prone conditions.

\textbf{DC-AI design principle.}
This benchmark instantiates a broader data-centric lesson: when a hand-crafted
proxy (IQR score) guides a sampler (MCMC chain), the proxy's coverage determines
the system's utility.
In tabular DC-AI settings---where quality defects span numeric, categorical, and
relational dimensions---no single numeric proxy captures the full quality surface.
The practical consequence is that the schema-free inductive bias of random uniform
sampling (every row is equally informative a priori) is better matched to the
actual structure of data quality defects than any proxy-guided alternative.
Profiling infrastructure for DC-AI pipelines should therefore be evaluated not
on theoretical elegance but on empirical accuracy across the data types it will
encounter---the approach this benchmark instantiates.

\textbf{Reproducibility.}
Our benchmark---covering six real datasets (up to 7.4M rows) and synthetic data
scaled to $5\times10^6$ rows, nine sampling strategies, and four error types---is
fully reproducible; code, data, and scripts are
available at \url{https://github.com/LaureBerti/progressive-profiling}
and permanently archived on Zenodo (DOI:
\href{https://doi.org/10.5281/zenodo.21628028}{10.5281/zenodo.21628028}).

\textbf{Future work.}
One concrete direction remains: resolving whether DAG-guided sampling helps on
genuinely structured relational or knowledge-graph data requires replacing the
random stub with a real dataset and running E2 at $n \geq 20$ seeds per $\rho$
level~\cite{button2013power}.
Stratified sampling with a proxy that captures \emph{categorical} quality defects
(not IQR-based) remains an open avenue.

\begin{credits}
\subsubsection{\ackname}
This work was carried out at IRD, ESPACE-DEV, Montpellier, France.

\subsubsection{\discintname}
The author has no competing interests to declare that are relevant to the content
of this article.
\end{credits}

\bibliographystyle{splncs04}
\bibliography{refs}

\end{document}